# Engineering zero modes, Fano resonance and Tamm surface states of 'bound states in the gapped continuum'

Ying Yang[1], and Yiming Pan[2]

1. Zhenjiang key laboratory for advanced sensing materials and devices and Department of Physics, Faculty of Science, Jiangsu University, Zhenjiang 212013, CHINA
2. Department of Physics of Complex Systems, Weizmann Institute of Science, Rehovot 76100, ISRAEL.

## Abstract

In this article, we developed a generalized coupled-mode theory for mixing an isolated state with the gapped continuum to unify the engineering mechanism of gap-protected zero modes, Fano resonance and Tamm surface states, even though those phenomena are diverse in numerous branches of physics and optics. By tuning the on-site potential and coupling strength of/with the isolated state, we found the operating characteristics from zero modes, Fano resonance and to Tamm modes, with explicitly demonstrating their localization, transmission spectra and distinct evolution dynamics of the 'bound states in the gapped continuum' (BIGC). As an extent, we designed two sandwich-like structures in the BIGC framework: (1) the 'isolated-continuous-isolated' (ICI), which leads to adiabatic elimination technique and topologically-protected edge states, and (2) the 'continuous-isolated-continuous' (CIC), which results into the domain wall dynamics. Finally, we proposed the optical waveguide array (WA) modelling of the modified Su-Schrieffer-Heeger dimerized lattice to achieve those interesting engineering processes, and meanwhile to shed light on the generalization of bound states in the continuum into photonics and condensed matters.

Humans love 'reinventing the wheel', also did nature. The coupled mode theory (CMT) is one of the well-established 'wheels', which is ubiquitous in numerous branches of physics, optics, mathematics and chemistry. The CMT has been proposed for many decades[1,2], even differs from formulations to formulations but its underlying assumption, based on the development of the solution to a perturbed or weakly interaction problem into a linear combination of its possible eigenmodes, exists as a principle for engineering in science and applications. For examples, the CMT has been widely utilized as a basic analytical formulation for different material systems such as optical waveguides and fibers [3,4], photonic crystals [5,6], semiconductors[7–9], and also applied to phenomena like Fano resonance[10–13], parity time (PT) symmetry[14–17], discrete solitons[18–20], Anderson localization[21–23], topological photonics[24–28].

Figure 1 depicted three typical setups in the CMT: the couplings between two isolated modes, two continuous modes and the hybrid continuous-isolated modes, respectively. In optics, the guide modes couple and transfer energy (or intensity) when the electric field amplitudes of two waveguides overlap (evanescent field), which indicates the fundamental typical properties of mode-coupling. The conventional field distribution between adjacent waveguides due to the superposition of two eigenmodes, with emergence of symmetric and antisymmetric modes as shown in Fig1a, whose dynamics relates to the chemical reaction process (bonding or anti-bonding). Fig.1b showed the gap generation in semiconductor physics as well as topological insulators. The gap generation classifies the phase transition from insulators and metals, and the topology of those emerging gaps can further enable to classify the insulator phases into topologically trivial or non-trivial families. More fundamentally, the gap generation as the typical demonstration offers the physical origin of mass terms in quantum field theories[29]. As a non-trivial exemplified, the gap generation is easily explained in the coupled-mode theory for two continuous bands crossing. Interestingly, Fig.1c showed the Fano resonance emerging from the bound states in the continuum (BIC), with the mixture of coupling the localized and continuous states altogether[30]. The physics of Fano resonance and its application have been widely investigated in numerous systems nowadays[31]. The microscopic origin of the Fano resonance arises from the constructive and destructive interference of a narrow discrete resonance with a broad spectral line or continuum, resulting in asymmetric profile of the localized state which was Lorentzian line shape before interference which also related to the generalized class of CMTs [11].

This work is to extend the coupling process of the isolated state with the 'gaped' continuum, as comparable to the conventional gapless case in Fig.1c. Owing to this inherent similarity between coupling in waveguides for optical field and hopping in lattice for electrons, waveguide array (WA) is chose to mimic Su-Schrieffer-Heeger (SSH) model[32] as the gapped continuum in this work as other researchers also did[33–35]. The gapped continuum of SSH model defines three regimes for the on-site potential engineering of the bound state in a unified picture: isolation within the gap (zero mode), resonance in the bulk (Fano resonance), and localization out of energy spectrum (Tamm mode). To describe the degeneracy of zero modes of SSH model and its domain-wall soliton in our extended CMT, two sandwich-like structures 'isolated-continuous-isolated' (ICI) and 'continuous-isolated-continuous' (CIC) are introduced. Both the adiabatic elimination (AE) and domain wall solitons have been intensively studied in SSH model[36], as also discussed in our sandwich-like structures. Via the tuning of the on-site potential of the localized state, the two brand-new AE technique and domain walls based on Tamm states engineering are found in the SSH dimerized model. The predictions in our extended CMT can lead to numerous application in photonics and physics, and may also be easily checked by the future observation on waveguide arrays.

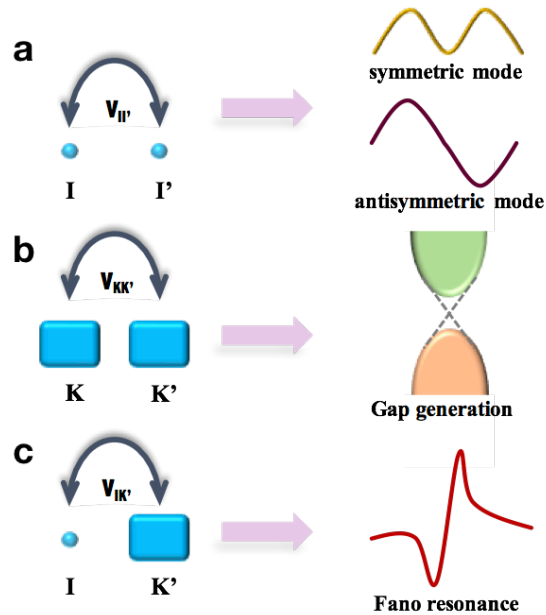

*Fig.1: The generalized coupled mode theory (CMT) for isolated-isolated states, continuous-continuous states, and mixed isolated-continuous states. (a) The symmetric and antisymmetric modes in waveguide systems, (b) gap generation in semiconductor physics, and (c) Fano resonance in scattering phenomena are unified in the CMT framework.*

## Modelling

Now, we construct the well-known Su-Schrieffer-Heeger (SSH) model[32,37] in our WAs as an example to define the gaped continuous modes. The corresponding dimerized Hamiltonian is given by

$$H_K = \sum_{n=1}^{2N}((\kappa_0 + (-1)^n \Delta\kappa)\,\varphi_n \varphi_{n-1}^\dagger + h.c.) + \sum_{n=1}^{2N} \beta_0 |\varphi_n|^2 \qquad (1)$$

where $\kappa_0$ is the average coupling strength ($2\kappa_0$ determines the band width $W = 2\kappa_0$ ), $\Delta\kappa$ is the staggered coupling strength and $\beta_0$ is the propagation constant related to the on-site potential in SSH lattice mode. Obviously, the energy dispersion is gapped where the gap $\Delta$ relates to $\Delta\kappa$. The gap generation $\Delta \neq 0$ of periodic lattice of 1D crystal is from Peierls' instability as claimed by Peierls' theorem[38]. $\varphi_n$ stands for the electromagnetic field amplitude in the $n^{th}$ waveguide and $2N$ is total number of waveguides in array. We then introduce an isolated waveguide which is described by

$$H_I = (\beta_0 + \Delta\beta)|\varphi_0|^2 \qquad (2)$$

where $\varphi_0$ is the field amplitude for isolated mode and $\Delta\beta$ defines the difference of propagation constant as compared to continuous mode. Thus, the coupling process of the mixture is given by

$$H_{int} = \int(V_{IK}^* \varphi_k^\dagger \varphi_0 + V_{IK} \varphi_0^\dagger \varphi_k)dk = V_0(\varphi_0^* \varphi_{2N} + h.c.) \qquad (3)$$

where $V_{IK}$ stands for the general coupling strength between isolated mode and the gaped continuous modes. Here, we simplified its coupling to be the nearest neighbor hopping ($V_0$) between the 2N$^{th}$ and 0$^{th}$ waveguide where the other hoppings are reduced to zero, and we noted that the non-nearest-neighbor interactions and configuration have also been widely studied in many literatures[11].

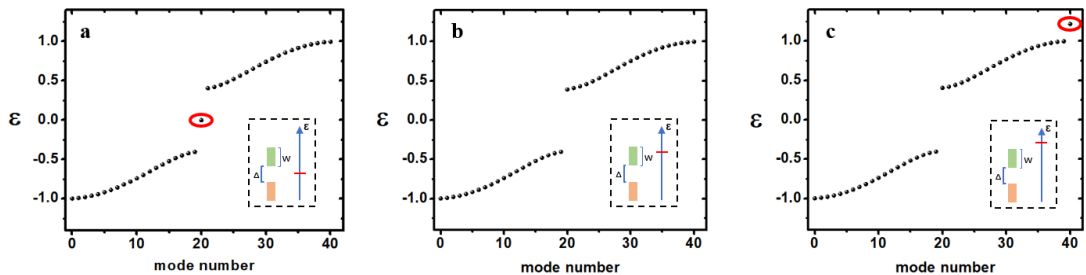

*Fig.2: The eigenmodes spectra for the mixing between the isolated mode (n=0) and the gapped continuous mode (2N=40). The isolated state lies (a) in the gap (<$\Delta$), (b) in the bulk (~W) and (c) out of the bulk spectra (>W), respectively.*

## Results

**The confinement of zero mode and Tamm mode** - Figure 2 showed the energy spectrum of our hybrid SSH model of zero modes, BIC and Tamm states by tuning the on-site potential as depicted in the insets. The simulation parameters are the average coupling strength $\kappa_0 = 0.5$, the staggered coupling $\Delta\kappa = 0.2$, the propagation constant of continuous mode is $\beta_0 = 1$ (all the other energy parameters are dimensionless by scaling with $\beta_0$), and total waveguide number is $2N + 1 = 41$, and the propagation constant of isolated mode is $\Delta\beta = 0, 0.6, 1.1$, respectively, and with the coupling between isolated and continuous modes $V_0$. Usually the SSH model has two distinct phases, one is topologically non-trivial but the other one is trivial. We take the trivial phase (B-phase) as the startup to eventually engineer the non-trivial phase (A-phase) in our extended CMT. In Fig. 2a, the isolated state localized in the gap (B-phase) ($\Delta\beta = 0 < \Delta$, $V_0 = 0.3$), which is protected by continuous mode, and is called the protected zero mode. For $\Delta\beta = 1.1, V_0 = 0.3$ in Fig. 2c, the on-site potential reaches out of the spectrum and generates the other kind of localized state, named Tamm surface state. Fig. 2b gives the BIC case with the on-site potential $\Delta\beta = 0.6$ and $V_0 = 0.3$, obviously the isolated state located in the continuous modes. Note that if the on-site potential $\Delta\beta$ approaches to 0.85 while the coupling between isolated and continuous modes is much weaker (e.g. $V_0 = 0.06$), Fano resonance would occur as expected (which will be discussed in next section).

From this eigenmode diagram, the differences between two localized modes are not obvious, we thus presented the field confinement of the protected zero mode and Tamm surface state in lattice space as shown in Fig. 3. The eigenstate of zero mode is protected by sublattice symmetry, and thus occupies only the even (or odd, depending on the count order) lattice. In our simulation, the light is launched into the $0^{th}$ waveguide, so the even waveguide number is occupied in Fig. 3a for zero mode and decays rapidly into the bulk. On the contrary, the Tamm surface state is decays exponentially into the bulk without the protection of such sublattice symmetry, as shown in Fig. 3b.

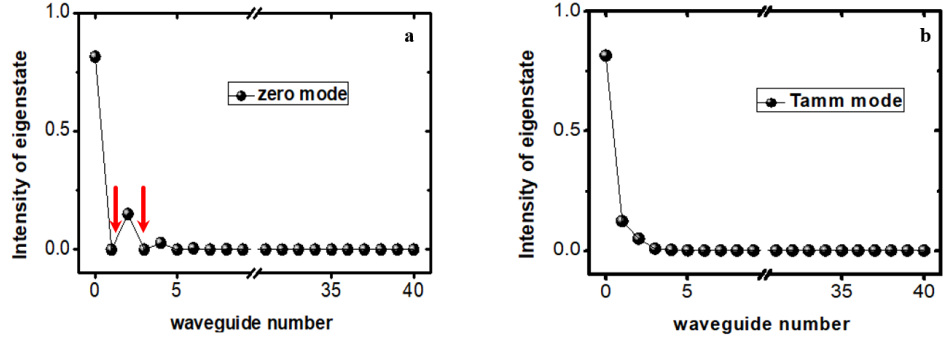

*Fig.3: The confinement configuration of eigenstates for (a) gap-protected zero mode and (b) Tamm surface state. The zero mode confines at the boundary and merely occupies the even waveguide number due to the protection of sublattice symmetry, but Tamm states decays into the bulk normally.*

Figure 4 showed the near-field propagation and its output for those three cases when the corresponding propagation constants of the isolated mode were set to $\Delta\beta = 0, 0.6, 1.1$ respectively. The incident field was launched from the $0^{th}$ waveguide as we participated to excite the zero mode and Tamm state. The field propagation retains its localization mostly on the $0^{th}$ waveguide but still few portion of field intensity was scattered into the continuous state due to the initial eigenstate projection. Meanwhile, Fig.4c and 4f showed the near-field dynamics and final output of Tamm states with similar confinement at the WA boundary. The difference of their confinement is that the output of zero mode still holds the sublattice symmetry and dismiss rapidly into bulk waveguides while that of Tamm mode decays exponentially as similar to the eigenstate configuration in Fig.3b.

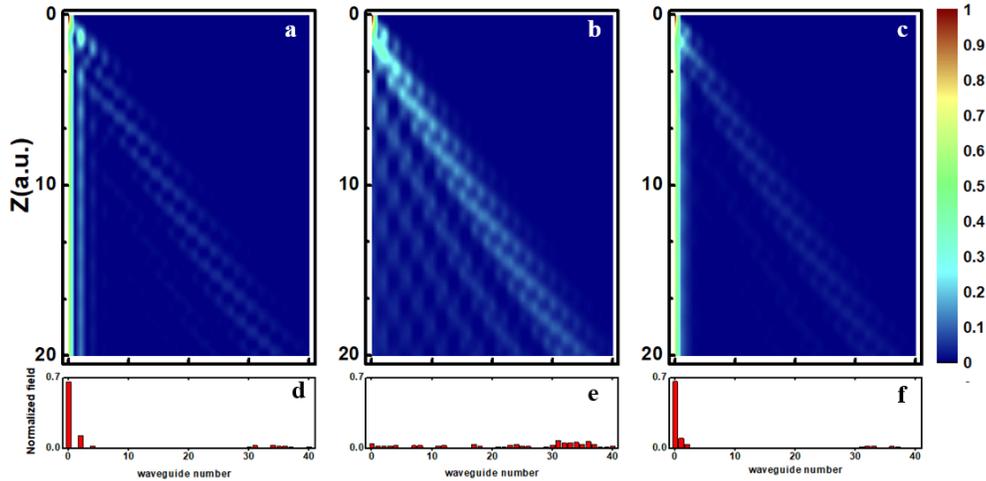

*Fig.4: The evolution dynamics and its output distributions in WA for zero mode excitation (a, d), bound state in the continuum (b, e) and Tamm mode excitation (c, f), respectively.*

**Fano resonance engineering between zero mode and Tamm state** - When the isolated mode laying in the bulk, the propagation pattern appears to be diffusive without typical resonant characteristics of diffraction management[3]. Figure 4b and 4e showed the spreading field evolution. However, we hardly distinguish its dynamical properties from the bulk in energy spectrum in Fig. 2b or field distribution in Fig. 4b. To demonstrate the Fano resonance, the transmission spectrum required to be calculated by transfer matrix method[39], and the results for different cases are thus given as shown in Figure 5. The simulation parameters are $\Delta\beta = 0.85$, and $V_0$ varied from 0, 0.06 (Fano resonance), 0.2 and 0.4. The spectrum shows that the isolated mode is strongly affected by the continuous mode, especially leading to Fano resonance around the same energy level of isolated mode. Figure 5 shows the typical antisymmetric Fano linewidth in transmission spectra at weak coupling V0=0.06 (red curve), this effect has been widely studied in WAs, atomic physics, and coupled quantum dots[11]. At $V_0 = 0$ (black curve), without coupling the isolated mode, then the transmission shows the bandwidth (one-branch) of continuous mode ranging from 0.4 to 1.0, where the zero mode and Tamm state are out of this range, with no contribution to the transmission spectrum.

By increasing the coupling strength $V_0$ (blue and cyan curves), we notice that the antisymmetric linewidth becomes broader and then disappears. Especially, the single transmission spectrum split into two sections at $V_0 = 0.4$. This indicates the hybridization of the isolated mode cut the continuous mode into two pieces, where one effective band starts from 0.4 to 0.85 and another one from 0.85 to 1.0. Because the two pieces of transmission spectrum are reduced to zero at energy ($\Delta\beta = 0.85$) of the isolated mode but still touched, it means that no gap emerges even though physically there are already two effective bands, which brings tiny changes in energy spectrum as shown in Fig. 5b. Figure 5d shows the field evolution with initial input from the 1$^{th}$ waveguide (not from the 0$^{th}$) The reason for this excitation is to create an interference loop of field scattering between the continuous mode and isolated mode where one path passes through the isolated state and reflected back into bulk state and another goes directly into the bulk, as demonstrated in Fig.5c. The interference thus leads to Fano resonance, whose transmission spectrum are shown in Fig.5a within different cases of coupling strengths.

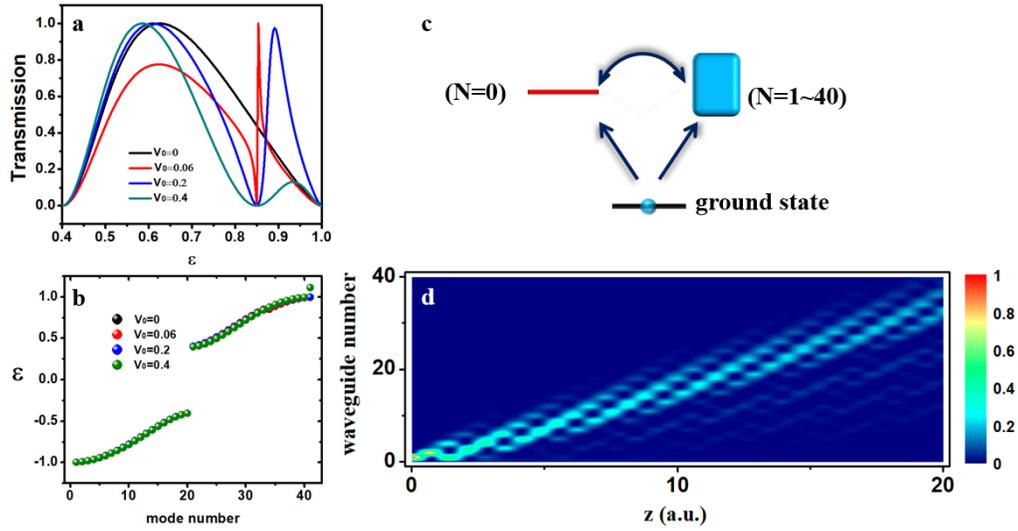

*Fig.5: Fano resonance is antisymmetric in the transmission spectrum when the isolated mode couples with the gapped continuum. (b) The energy level of isolated state is $\Delta\beta = 0.85$ which lies in the range of bulk band from 0.4 to 1.0. The black curve for the uncoupled case ($V_0$=0). (c) The interference origin of Fano resonance and (d) the evolution dynamics of Fano resonance when the 1$^{st}$ waveguide is excited at the input.*

## Discussion

**Adiabatic Elimination in the "isolated-continuum-isolated" setup** - Adiabatic Elimination (AE) is a powerful decomposition technique in quantum optics and atomic physics that allow one to eliminate certain degrees of freedom out of the dynamics under investigation[36,40]. The AE decomposition is helpful to deal with complicated multilevel system and offers simplification of the systematic Hilbert space into the effect restricted subspace, thus has been widely applied and intensively studied for many decades. For our concern, we can construct a sandwich-like structure based on our generalized CMT that is setup as "isolated-continuum-isolated" configuration. By tuning the mixing coupling between isolated and continuous modes separately, we achieved the effective coupling between two isolated modes by integrating out the continuous bulk state irrelevantly. With extending our CMT, the integration procedure requires approximation to the second-order, which can be derived from the perturbation methods such as Wolf-Schrieffer transformation, Green function and advanced path integral approaches[36,41]. Thus, we eventually obtained the effective hybridization of protected zero modes and Tamm states in the standard procedure of adiabatic elimination, which can be checked in further related experiments.

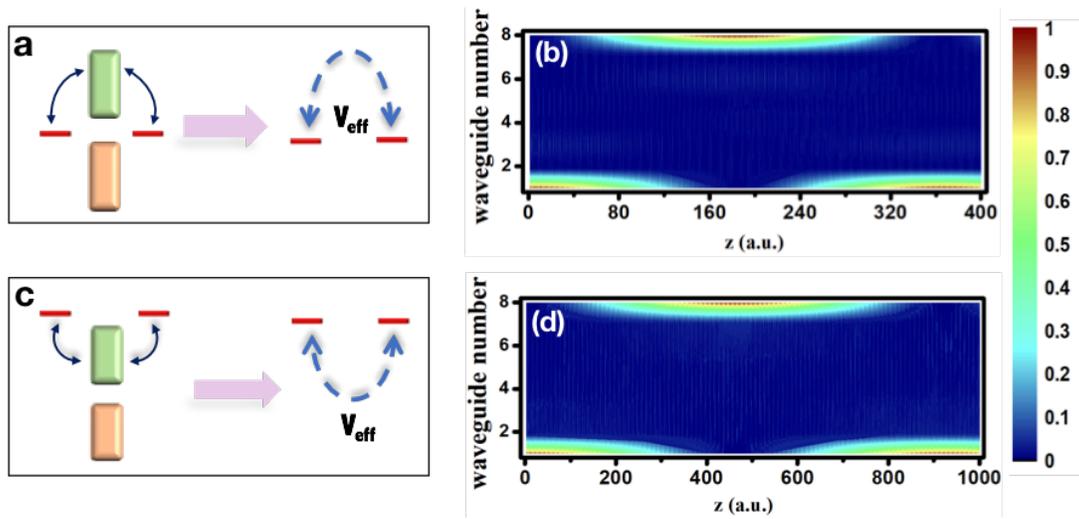

*Fig.6: The configuration of "isolated-continuum-isolated" (ICI) sandwich-like structure. The adiabatic elimination of the coupled zero modes (a-b), and of the coupled Tamm modes (c-d), respectively.*

Figure 6b and 6d show the field evolution of AE in our sandwich setup. The parameters are $2N + 2 = 8, V_1 = V_8 = 0.1$, and $\Delta\beta = 0.1$ in Fig. 6b, and $\Delta\beta = 1.1$ in Fig. 6d. The field propagates along with $1^{th}$ waveguide and spreads into bulk waveguides and thus recovers from the other end waveguide. The inner waveguides play the intermediate transition role to effectively couple two end states which holds no strongly fields in propagation and thus enable to be eliminated out. Also, the similar AE pattern is found for the Tamm states as shown in Fig.6c and d, in which the isolated propagation constant is tuned to exceed the bulk spectrum. In addition, the AE for zero modes can be explained by the finite size effect of topological phase (A phase) of SSH modelling. The two degenerated zero modes exists by protection of sublattice symmetry. When the total waveguide number of the array reduces to 2N=8, the zero modes overlap in the bulk and lead to the hybridization and energy split. The effective coupling strength from hybridization is proportional to the decay factor $e^{-\gamma N}$, where the coefficient $\gamma$ relates to its gap. The hybridization of Tamm states is also founded in our sandwich structure, which shows the advantage of our CMT. Note that the Fano-type BIC setup (when $\Delta\beta = 0.85$) is a special subtle case related to coupling strengths in our framework. There is then no AE when the adiabatic condition is violated if the coupling is strong (the energy of isolated state approaches to the top of the upper band) and isolated state is much more easily to spread into continuum. However, as the coupling is decreasing, AE still appears since the isolated state is more like the BIC case in weak coupling regime.

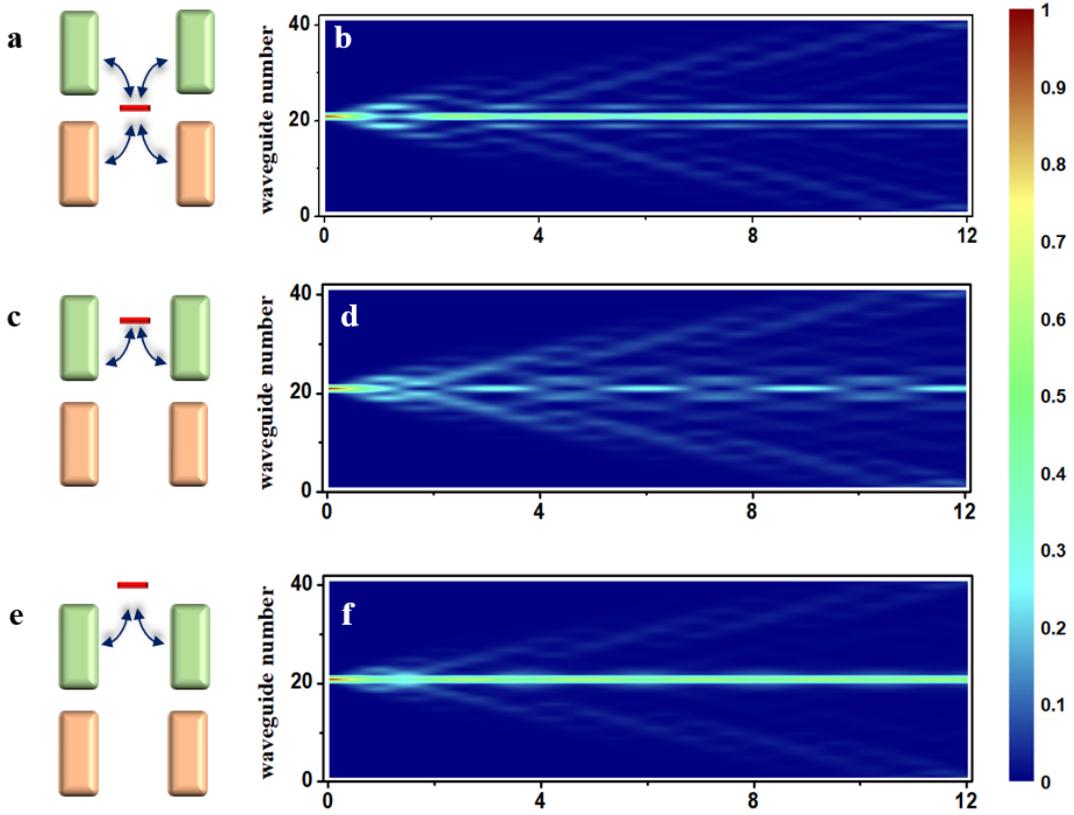

*Fig.7: The configuration of "continuum-isolated-continuum" (CIC) domain wall structure. The solitary dynamics for zero modes (a-b), BIC (c-d) and Tamm modes (e-f), respectively.*

**Domain Walls in the "continuum-isolated-continuum" setup** - The second application is in fact the other kind of sandwich-like structure as the "continuum-isolated-continuum" (CIC) configuration, in other words, as the domain walls (DWs). Figure 7a, c and e show three kinds tunable DWs. At $\Delta\beta = 0$ where the isolated mode lies in the gap, the protection is enhanced by by-side continuums and give rise to the topological soliton, as first proposed in Polyacetylene (1979)[32]. In the context of TIs, "continuum-isolated" configuration is viewed altogether as topological phase and create an interference with the normal insulator (B-phase). The band continuity requires the gap to "open-closed-reopen" process topologically when phase passing through the interference[42]. The 'closed' process of gap generation leads to the soliton excitation which is called as "bulk-edge correspondence" (BEC)[42,43], and the topologically-protected interference between TI and normal insulator is called as domain walls. The propagation pattern of the protected zero modes DW is shown in Fig. 7b, with same parameter as set in Fig.4b. Similarly, Fig. 7e and f show the evolution of Tamm modes DW. We have to note here that the protection by gap generation actually have no contribution to its confinement of Tamm modes DW.

Conversely, for the BIC setup, the DW disappear without surprise as shown in Fig. 7c-d and has already been explained in Fig. 4b.

We would like to address some extensions on the underlying dimensions of the bound state and the continuum in the generalized coupled mode theory. In our theory, we can actually view the dimension of isolated mode as zero-dimension (0D) as compared to that of continuous mode as one-dimension (1D), which relates one single free parameter. Consider the dimensionality of coupled mode, we can easily extend our theory into the mixture between (N-1)-dimensional 'isolated' modes and N-dimensional 'continuous' modes in Hilbert space. If the isolated mode (the dimension is usually lesser than N dimension) lies in the gap of continuous mode (N), thus the immunity by protecting from the scattering into the bulk is also viewed as bulk-edge correspondence in principle. The general BEC is also viewed as bulk-vortex correspondence (the coupling between 3D and 1D), or bulk-singularity correspondence (the coupling between 3D and 0D, also topological defects), which depends on the specific topology configuration. The capability of the protection is critically defined by the scale of gap compare to the strength of mixed coupling between isolated mode and continuous mode.)

## Conclusions

The coupled mode theories (CMTs) have many versions of presence in plenty of branches and disciplines of science and have already been explored for many years. Here, we developed a generalized CMT for the mixture of isolated states and the gapped continuum. In our extended CMT framework with tuning the on-site potential or the coupling strength of the isolated state, we reported a simplest demonstration to achieve the protected zero modes, Fano resonance and Tamm surface states consistently. Furthermore, by engineering the extended CMT into the sandwich-like structures, we realized the adiabatic elimination and domain walls dynamics in two different setups, especially for the realization of novel distinguished hybridization and domain wall of Tamm modes. Thus, the protection of zero modes, transmission of Fano resonance, confinement of Tamm states, adiabatic elimination technique and physics of domain walls are elegantly unified in our extended CMT of 'bound states in the gapped continuum'(BIGC), which enables to shed light on the various applications in many fields relating to mixing between isolated and continuous modes.

## Acknowledgement

This work was supported by Talent Fund of Jiangsu University (17JDG014), by DIP—German-Israeli Project Cooperation (7123560301), by the BSF-NSF (2014719), by Icore— Israel Center of Research Excellence program of the ISF, and by the Crown Photonics Center. Correspondance and requests for materials should be addressed to Y. P. (yiming.pan@weizmann.ac.il) or Y.Y (yingyang@ujs.edu.cn).